\documentclass[aps,prl,times,twocolumn,superscriptaddress,showpacs,a4paper]{revtex4-1}

\usepackage{colortbl}
\usepackage{amsfonts,amsmath,amssymb}
\usepackage[dvips]{graphicx}
\usepackage{hyperref,tikz}
\usepackage{tabularx,hhline}
\usepackage{float} 
\usepackage{dcolumn,stmaryrd}

\begin{document}

\title{Are multiphase competition \& order-by-disorder the keys to understanding Yb$_{2}$Ti$_{2}$O$_{7}$?}

\author{L.D.C. Jaubert}
\affiliation{Okinawa Institute of Science and Technology Graduate University,
Onna-son, Okinawa 904-0395, Japan}

\author{Owen Benton}
\affiliation{Okinawa Institute of Science and Technology Graduate University,
Onna-son, Okinawa 904-0395, Japan}

\author{Jeffrey G. Rau}
\affiliation{Department of Physics and Astronomy, University of Waterloo, 200 University Avenue West, Waterloo, Ontario, N2L 3G1, Canada}

\author{J. Oitmaa}
\affiliation{School of Physics, The University of New South Wales, Sydney 2052, Australia}

\author{R.R.P. Singh}
\affiliation{Department of Physics, University of California, Davis, California 95616, USA}

\author{Nic Shannon}
\affiliation{Okinawa Institute of Science and Technology Graduate University,
Onna-son, Okinawa 904-0395, Japan}

\author{Michel J.P. Gingras}
\affiliation{Department of Physics and Astronomy, University of Waterloo, 200 University Avenue West, Waterloo, Ontario, N2L 3G1, Canada}
\affiliation{Perimeter Institute for Theoretical Physics, 31 Caroline North, Waterloo, Ontario, N2L 2Y5, Canada}
\affiliation{Canadian Institute for Advanced Research, Toronto, Ontario, M5G 1Z8, Canada}

\date{\today}
\begin{abstract}
If magnetic frustration is most commonly known for undermining long-range order, as famously illustrated by spin liquids, the ability of matter to develop new collective mechanisms in order to fight frustration is no less fascinating, providing an avenue for the exploration and discovery of unconventional properties of matter. Here we study an ideal minimal model of such mechanisms which, incidentally, pertains to the perplexing quantum spin ice candidate Yb$_{2}$Ti$_{2}$O$_{7}$. Specifically, we explain how thermal and quantum fluctuations, optimized by order-by-disorder selection, conspire to expand the stability region of an accidentally degenerate continuous symmetry U(1) manifold against the classical splayed ferromagnetic ground state that is displayed by the sister compound Yb$_{2}$Sn$_{2}$O$_{7}$. The resulting competition gives rise to multiple phase transitions, in striking similitude with recent experiments on Yb$_{2}$Ti$_{2}$O$_{7}$ [Lhotel \textit{et al.}, Phys. Rev. B \textbf{89} 224419 (2014)]. Considering the effective Hamiltonian determined for Yb$_{2}$Ti$_{2}$O$_{7}$,  we provide, by combining a gamut of numerical techniques, compelling evidence that such multiphase competition is the long-sought missing key to understanding the intrinsic properties of this material. As a corollary, our work offers a pertinent illustration of the influence of chemical pressure in rare-earth pyrochlores.
\end{abstract}
\pacs{}
\maketitle


The wide interest in magnetic frustration largely stems from the diversity of unconventional phenomena it begets, ranging from anomalous Hall effect~\cite{Taguchi01a} to multiferroicity~\cite{Mostovoy06a}, to name only a few. The engine for this diversity is the inherent indecisiveness of frustrated magnets towards ordering, opening an avenue for new mechanisms to control their low-temperature properties. The understanding of such mechanisms not only enlarges our broad understanding of the principles via which matter organizes itself; it is also the necessary first step in order to ultimately control the exotic properties of frustrated magnets.

In the context of emergent degeneracies in pyrochlores~\cite{Yan13a,Canals08a}, we present here a thorough analysis of the ordering mechanisms of a realistic microscopic model of multiphase competition. We believe our theory is the missing key to unlock many of the puzzling features of Yb$_{2}$Ti$_{2}$O$_{7}$, whose ordering $-$ or absence of $-$ has been a matter of debate for nearly 15 years~\cite{Hodges02a,Yasui03a,Ross09a,Yaouanc11a,Ortenzio13a}, discussed in the context of quantum spin liquid~\cite{Ross11a,Savary12a,Savary13a}, Higgs mechanism~\cite{Chang12a} and magnetic monopoles~\cite{Pan14a,Pan15a}. Specifically, our theory accounts for the multi-step ordering process and field dependence of Yb$_{2}$Ti$_{2}$O$_{7}$~\cite{Lhotel14a} while providing a natural setting to understand its sample dependence issue~\cite{Yaouanc11a,Ross12a}.

\begin{figure}[h!]
\centering\includegraphics[width=\columnwidth]{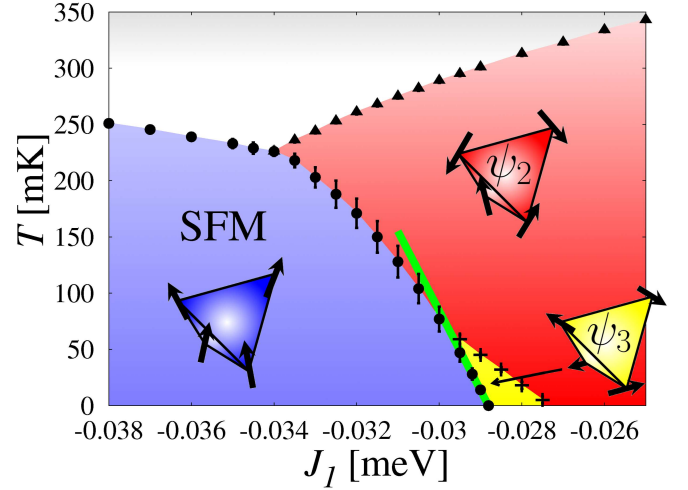}
\caption{
Multiphase competition of the anisotropic nearest-neighbor pyrochlore model of Eq.~(\ref{eq:ham}), as determined by classical Monte Carlo simulations for $\{J_{i=2,3,4}\}=\{-0.22,-0.29,0\}$ meV. The $\psi_{2}$ and $\psi_{3}$ phases are selected by order-by-disorder within the antiferromagnetic U(1) manifold and separated from the splayed ferromagnet (SFM) by a first-order transition, whose slope agrees with classical low-temperature expansion (green line, see Appendices). Rotating all spins of a given $\psi_{2}$ state by the same angle around their local easy-axes produces the entire U(1) manifold, including the $\psi_{3}$ states~\cite{Poole07a}. The ``splaying'' angle of the spins in the SFM ground state is uniquely determined by the $\{J_{i}\}$ parameters.
}
\label{fig:PDT}
\end{figure}

Starting from the most general model for anisotropic nearest-neighbor exchange on the pyrochlore lattice~\cite{Curnoe07a,Curnoe08a}, we consider a range of parameters known to describe Yb$_2$Ti$_2$O$_7$ [\onlinecite{Ross11a}]. We show that a splayed ferromagnetic (SFM) phase such as the one observed in the sister compound Yb$_{2}$Sn$_{2}$O$_{7}$~\cite{Yaouanc13a,Dun13a} competes against a degenerate U(1) manifold for ordering (see Fig.~\ref{fig:PDT}). As in Er$_{2}$Ti$_{2}$O$_{7}$, a strong contender for experimental realisation of order-by-disorder (ObD)~\cite{Champion04a,Savary12b,Zhitomirsky12a,Wong13a,Oitmaa13a}, soft modes of excitations lift the U(1) degeneracy. This degeneracy lifting creates an additional competition between the so-called $\psi_{2}$ and $\psi_{3}$ states within the U(1) manifold. Using a wide range of techniques, we find that both thermal and quantum fluctuations enhance the stability of the U(1) manifold to the detriment of the SFM phase. As a consequence, quantum fluctuations bring Yb$_2$Ti$_2$O$_7$ close to the boundary between all three phases where multiple phase transitions take place. In addition to the theoretical interest of understanding how multiple ObD selections take place within a classically energetically unstable manifold~\cite{Pinettes02a,Yan13a,Mcclarty14b}, our work offers a natural explanation of the properties of Yb$_2$Ti$_2$O$_7$, as recently observed in bulk measurements~\cite{Lhotel14a}.\\


\textit{Model -- } We consider the anisotropic Hamiltonian
\begin{eqnarray}
\mathcal{H} = \sum_{\langle ij\rangle} \vec S_{i} \;\bar{\mathcal{J}}_{ij}\; \vec S_{j}\quad{\rm with}\quad 
\bar{\mathcal{J}}=
\begin{pmatrix}
J_{2} & J_{4} & J_{4}\\
-J_{4} & J_{1} & J_{3}\\
-J_{4} & J_{3} & J_{1}
\end{pmatrix}
\label{eq:ham}
\end{eqnarray}
where $i,j$ are pyrochlore nearest-neighbors. All coupling matrices $\bar{\mathcal{J}}_{ij}$ can be deduced from $\bar{\mathcal{J}}$ by appropriate symmetry transformations~\cite{Ross11a}, with only four independent parameters $\{J_{i=1..4}\}$ allowed by the symmetry of the pyrochlore lattice~\cite{Curnoe07a,Curnoe08a}. Our work focuses on the parameter line $J_{1}\in[-0.09:0]$ meV and $\{J_{i=2,3,4}\}=\{-0.22,-0.29,0\}$ meV, relevant to Yb$_{2}$Ti$_{2}$O$_{7}$ for $J_{1}=-0.09(3)$\footnote{the number between parentheses represents the error bar on the last digit, \textit{i.e.} $J_{1}=-0.09\pm0.03$ here.}~\cite{Ross11a} and including the $T=0$ boundary between SFM and U(1) phases at $J_{1}=-0.029$ (see Fig.~\ref{fig:PDT}). We consider both quantum spins $S=1/2$ and classical Heisenberg spins ($|\vec S|=1/2$) whose phase diagrams have been studied in~Refs.~[\onlinecite{Savary12a,Lee12a,Savary13a,Wong13a,Hao14a}] and~[\onlinecite{Yan13a}], respectively.\\

\newcolumntype{C}{>{$}c<{$}}
\newcolumntype{L}{>{\centering $} p{1.7cm}<{$}}
\newcolumntype{S}{>{\centering $} p{1.2cm}<{$}}
\renewcommand{\arraystretch}{2}
\begin{table}[ht]
\begin{center}
\begin{tabular}{||C|C|C|C|C|C||}
\hhline{|t:======:t|}
& \textrm{Classical} & \multicolumn{4}{c||}{\textrm{Quantum}}\\
\hhline{||~-----||}
& \textrm{MC} & \textrm{LSW} & \textrm{ED} & \textrm{NLC} & \textrm{HTE} \\
\hhline{||------||}
T>0 & -0.0340(5) & n/a & n/a & -0.070(3) & -0.06(3)\\
T=0 & -0.0289(1) & -0.062 & -0.064(2) & n/a & n/a \\
\hhline{|b:======:b|}
\end{tabular}
\end{center}
\caption{
Critical value of the exchange parameter $J_{1}$ [meV], separating the splayed ferromagnet (SFM) from the U(1) manifold, as estimated from different methods (Monte Carlo, linear spin-wave, exact diagonalization, numerical linked-cluster and high-temperature expansion) at zero temperature and upon cooling from high temperature. Quantum and thermal fluctuations work together to stabilize the U(1) manifold over the SFM phase. In particular, quantum effects bring the SFM/U(1) boundary within error bars of Yb$_{2}$Ti$_{2}$O$_{7}$ parameters~\cite{Ross11a}.
}
\label{table:PD}
\end{table}

\textit{Classical thermal fluctuations --} Starting at $T=0$, the 6-fold degenerate SFM phase persists for $J_{1}<-0.029$ before giving way to the one-dimensional degenerate U(1) manifold for $J_{1}>-0.029$. At the boundary, new continuously degenerate ground states emerge which confer additional zero-mode fluctuations to $\psi_{3}$ configurations~\cite{Yan13a}.

This zero-temperature framework sets the scene for the phase diagram of Fig.~\ref{fig:PDT} computed by Monte Carlo (MC) simulations using parallel tempering~\cite{swendsen86a,geyer91a} and over-relaxation~\cite{creutz87a}. The U(1) degeneracy does not survive thermal fluctuations and collapses predominantly in the $\psi_{2}$ configurations, with a small $\psi_{3}$ island around the boundary due to the above-mentioned soft modes of excitations. An interesting consequence of this ObD competition is that the $\psi_{2}/\psi_{3}$ phases are precisely selected to optimize the entropy of the U(1) manifold for a given set of parameters and temperature. At finite temperature, this optimization puts the energetically selected SFM phase at a disadvantage and gives rise to multiple phase transitions for $J_{1}\in[-0.034:-0.029]$. Since the Hamiltonian of Eq.~(\ref{eq:ham}) supports a large variety of emergent degeneracies and potential ObD transitions at the boundaries between ordered phases~\cite{Yan13a,Benton14a}, we expect such a phenomenology to be a common feature of pyrochlores~\cite{Chern10b,Mcclarty14b} and frustrated magnetism in general~\cite{Pinettes02a}. But since temperature is not the only source of fluctuations, how do quantum fluctuations fit in this picture ?\\


\textit{Quantum fluctuations at zero temperature --} Since we know the competing classical phases (SFM, $\psi_{2}$ and $\psi_{3}$), it is of interest to analyze their stability in the semiclassical limit using linear spin-wave theory (LSW)~\cite{Fazekas99a}. However, when applied to classically \textit{unstable} states, LSW usually becomes rather tedious as it requires higher order terms in the spin wave expansion. To circumvent this problem, we used the method developed in Ref.~[\onlinecite{Coletta13a}], which provides an upper bound of the semiclassical $\psi_{2}/\psi_{3}$ energies for $J_{1}<-0.029$ meV (see Appendix). Keeping in mind that this approach underestimates the stability of the U(1) manifold, LSW shows that the semiclassical $T=0$ frontier is shifted by quantum zero-point fluctuations from $-0.029$ meV down to $-0.062$ meV (see Table~\ref{table:PD}).

We now consider the full quantum nature of the spins. Since frustration is already at play in the constituting bricks of the pyrochlore lattice, namely the tetrahedra, exact diagonalization of a finite number of tetrahedra provides a good indication of the local influence of quantum effects. To preserve the symmetry of the pyrochlore lattice, we consider clusters of 4 and 16 spins, forming respectively 1 and 5 tetrahedra and allowing for standard ED calculations. Defining the order parameter $M$ and correlator $C=\langle M^{2}\rangle-\langle M\rangle^{2}$ of a given phase, then the quantity $\Delta C=C_{\rm U(1)}-C_{\rm SFM}$ is a direct measure of the SFM/U(1) competition, bringing the $T=0$ frontier to $J_{1}\approx-0.052(2)$ and $-0.064(2)$ for $N=4$ and $N=16$ respectively, in agreement with the semi-classical results (see Table~\ref{table:PD}).\\


\textit{Quantum fluctuations at finite temperature --} Even if the joint analysis of thermal and quantum fluctuations is particularly challenging for such a frustrated problem, the build up of correlations when approaching the phase transition from high temperature remains accessible thanks to numerical linked-cluster computation~\cite{Rigol07b,Applegate12a,Tang13a} and high-temperature expansion~\cite{Oitmaa06a} (see Appendix for technical details).

\begin{figure}[ht]
\centering\includegraphics[width=\columnwidth]{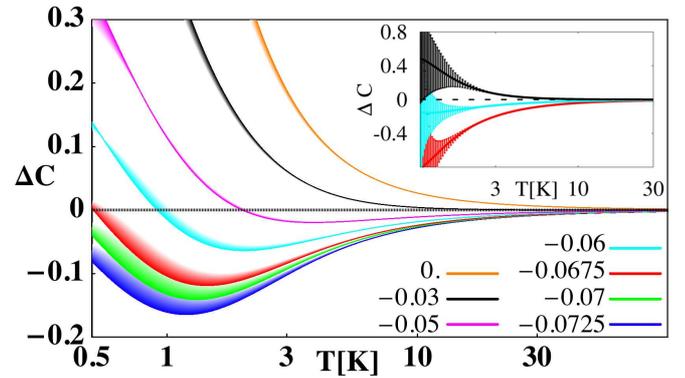}
\caption{
The difference of correlators $\Delta C=C_{\rm U(1)}-C_{\rm SFM}$ computed with NLC confirms the quantum shifting of the boundary towards more negative values of $J_{1}$ than for the classical system, estimated at $J_{1}=-0.070(3)$ meV. $J_{1}$ is given in the caption while $\{J_{i=2,3,4}\}=\{-0.22,-0.29,0\}$. \textit{Inset:} $\Delta C$ as computed from HTE for $J_{1}=-0.09$ (red), $-0.06$ (cyan) and $-0.03$ (black). The breadth of each curve represents the uncertainty for NLC and HTE at low temperature.
}
\label{fig:corr}
\end{figure}

HTE confirms the shifting of the boundary down to $J_{1}=-0.06(3)$ meV and with transition temperatures lower than 500 mK (see inset of Fig.~\ref{fig:corr}), in agreement with our classical simulations. As for NLC, at high temperature where quantum effects ultimately disappear, $\Delta C$ changes sign at the \textit{classical} limit $J_{1}\approx-0.03$ meV, which can be understood from $\beta^{2}$ terms in high-temperature expansion. Then, as temperature decreases for $J_{1}< -0.03$, instead of diverging towards SFM ordering, $\Delta C$ shows a clear upturn towards enhanced U(1) correlations (see Fig.~\ref{fig:corr}). This upturn is adiabatically evolving to lower temperature as $J_{1}$ is decreased. NLC thus puts the U(1)/SFM frontier at $J_{1}=-0.070(3)$ meV and, together with HTE, confirms the quantum nature of the boundary shift observed in ED and LSW, but now observed at non-zero temperature (see Table~\ref{table:PD}).\\


\textit{Multiphase competition in Yb$_{2}$Ti$_{2}$O$_{7}$ --} The Hamiltonian~(\ref{eq:ham}) was shown to fit well inelastic neutron scattering data of Yb$_{2}$Ti$_{2}$O$_{7}$ under large field~\cite{Ross11a}, allowing for the estimation of the exchange parameters $\{J_{i=1,2,3,4}\}=\{-0.09(3),-0.22(3),-0.29(2),0.01(2)\}$ meV. This set of parameters has been useful to understand the paramagnetic and high field regimes~\cite{Ross11a,Applegate12a,Chang12a,Savary13a,Hayre13a,Yan13a}. But many questions remain open at low temperature and zero field. For instance, while single crystals show a broad peak at $\approx 185$ mK in specific heat, powder samples typically display a sharp peak at $\approx 265$ mK~\cite{Yaouanc11a,Ross12a,Ortenzio13a}, suggesting the influence of structural disorder as in Tb$_{2}$Ti$_{2}$O$_{7}$~\cite{Taniguchi13a} or in Dy$_{2}$Ti$_{2}$O$_{7}$~\cite{Sala14a}. But interestingly, recent bulk measurements have brought to light new generic features between powder and single crystals~\cite{Lhotel14a}.

\begin{figure*}[ht]
\centering\includegraphics[height=6.5cm]{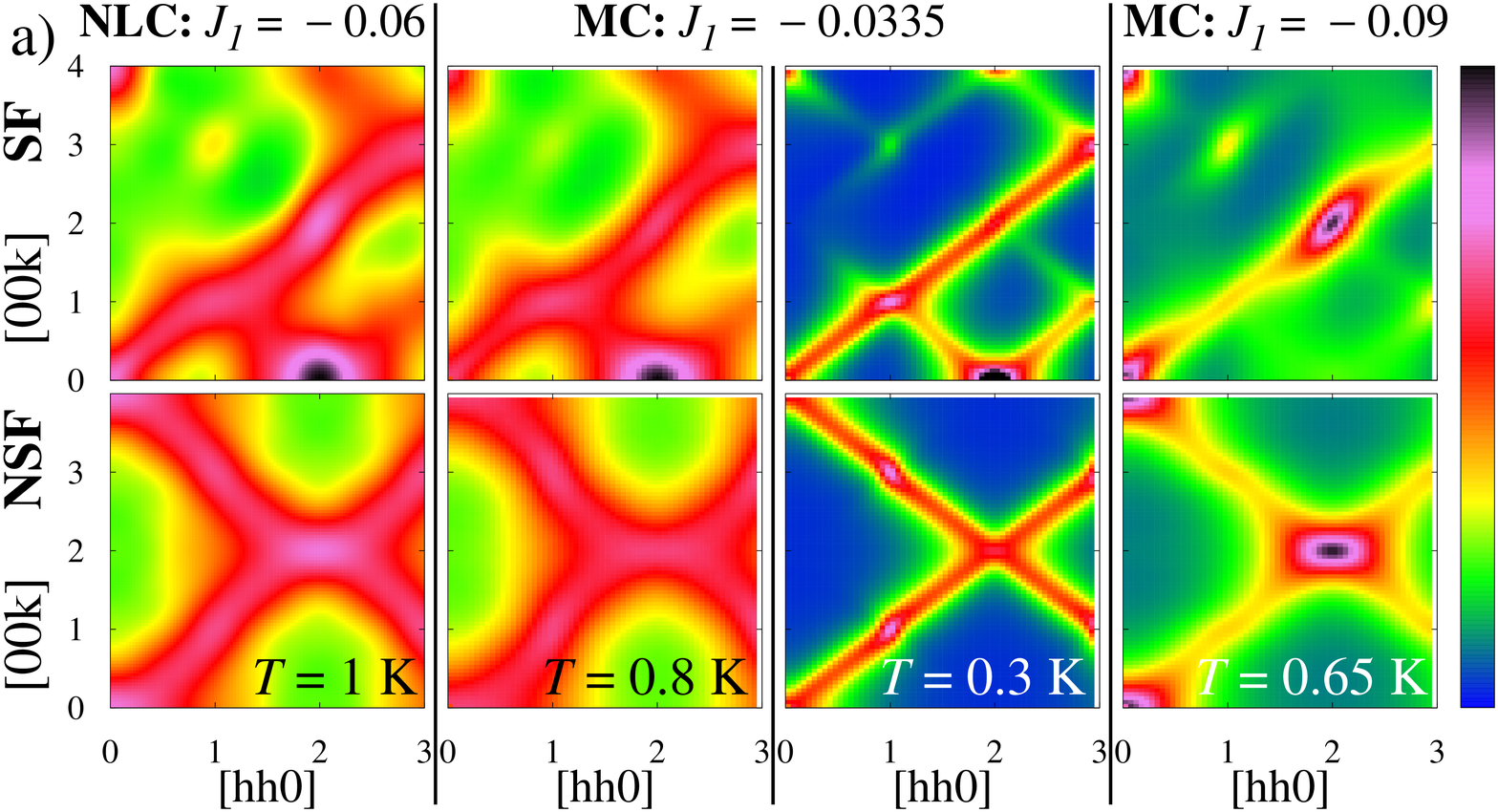}
\hspace{0.5cm}
\centering\includegraphics[height=6.cm]{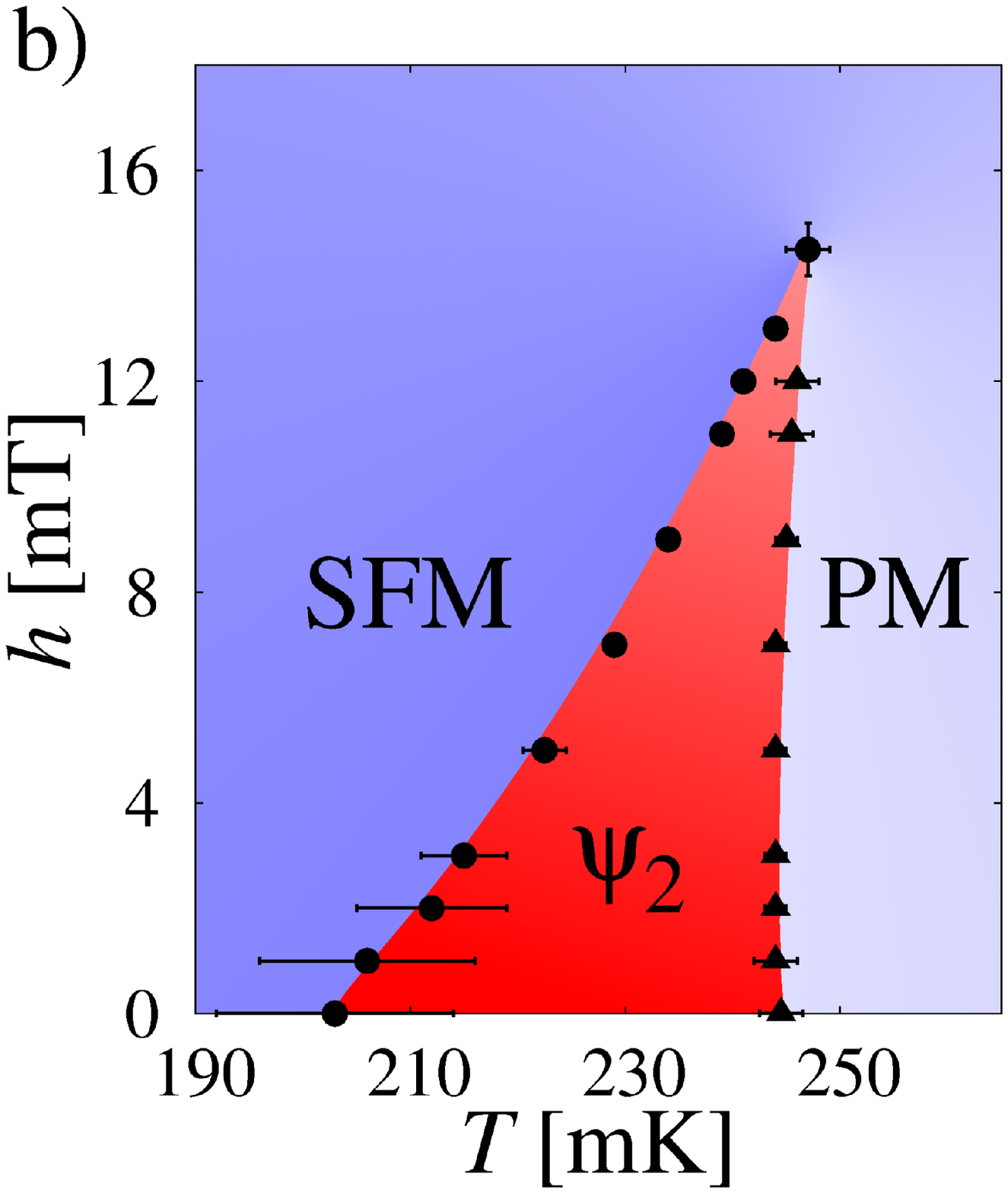}
\caption{
\textit{Application to Yb$_{2}$Ti$_{2}$O$_{7}$:} a) structure factor as measured by neutron scattering and b) phase diagram in a field. 
a) The spin-flip (SF, top panels) and non-spin-flip (NSF, bottom panels) calculated by quantum NLC and classical Monte Carlo are almost identical when $J_{1}$ is shifted from $-0.0335$ meV to $-0.06$ meV (please note that the temperature has been renormalized by 5/4 for a better agreement). This agreement confirms the quantum shifting of the boundary at finite temperature. When approaching the phase transition, the MC structure factor for $J_{1}=-0.0335$ meV reproduces the characteristic features of Yb$_{2}$Ti$_{2}$O$_{7}$ neutron scattering data measured at 300 mK (to be compared with Fig.~2a of Ref.~[\onlinecite{Chang12a}]). On the other hand, the comparison between experiments~\cite{Chang12a} and classical simulations using the parametrization of Ref.~[\onlinecite{Ross11a}] ($J_{1}=-0.09$ meV) are noticeably less successful, especially around (220)~\cite{Yan13a}. This disagreement is \textit{not} a criticism of the parametrization by Ross \textit{et al.}, but rather an emphasis on the importance of quantum fluctuations. The temperature of 0.65 K in the right panels has been chosen such that the ratio between measurement and transition temperatures is the same as in Ref.~[\onlinecite{Chang12a}]. The color scale is fixed from 0 to the maximum SF intensity, except for the panel at 0.3 K where the color scale was chosen for visual comparison with experiments~\cite{Chang12a}. 
b) When a field is applied along the [001] direction, the antiferromagnetic $\psi_{2}$ phase gradually disappears in MC simulations. Our theory thus explains why the first-order transition only persists at low field in Yb$_{2}$Ti$_{2}$O$_{7}$~\cite{Lhotel14a}. We used $J_{1}=-0.033$ meV and the anisotropic g-tensor of Yb$_{2}$Ti$_{2}$O$_{7}$~[\onlinecite{Hodges01a}]: $g_{\perp}=4.18$ and $g_{\sslash}=1.77$. 
}
\label{fig:SQ}
\end{figure*}

Based on these latter experiments~\cite{Lhotel14a}, and thanks to the present analysis, we believe it is finally possible to flesh out a common framework for the powder and single crystals which magnetically order. In Ref.~\cite{Lhotel14a}, both samples undergo a \textit{double phase transition} with a high-temperature non-ferromagnetic transition followed by a low-temperature ferromagnetic first-order transition. Furthermore, the application of a magnetic field $h$ does not destroy the lower-temperature ferromagnetic transition, but rather increases its temperature for $h> 5$ mT; the transition remaining first order up to $h_{c}\approx 20$ mT, before becoming continuous or vanishing, which is experimentally difficult to distinguish. These experimental results fit perfectly within the framework of our theory. The double transition is a direct consequence of the SFM/U(1) competition, as shown in Fig.~\ref{fig:PDT} and for a similar range of temperature as in experiments. Furthermore, since the U(1) manifold is antiferromagnetic, it does not couple with $h$. The magnetic field thus only favors the SFM phase, and the first-order ferromagnetic transition is expected to persist until the U(1) phase is destroyed at $h_{c}$. Monte Carlo simulations confirm this scenario with $h_{c}\approx 15$ mT, followed by a crossover for $h>h_{c}$ (see Fig.~\ref{fig:SQ}.b). We also mention as a further similarity the experimental presence of a reversible bump in the single crystal magnetisation at $\approx 180$ mK, \textit{between} the two transitions. According to our MC simulations, such feature could correspond to a $\psi_{2}/\psi_{3}$ ObD transition, but with the caveat that the $\psi_{3}$ phase does not persist above 50 mK in our classical phase diagram. In that case, structural disorder may play an important role~\cite{Ross12a,Taniguchi13a,Sala14a}, since it is known to i) favor $\psi_{3}$ over $\psi_{2}$ order~\cite{Maryasin14a} and ii) to be stronger in single crystals than powder samples where no bump was observed~\cite{Lhotel14a}.

It should be noted that even if the parametrization that we used, taken from Ref.~[\onlinecite{Ross11a}], was done on samples different from the ones in Ref.~[\onlinecite{Lhotel14a}], the quantum shifting of the SFM/U(1) boundary would bring this classical scenario within the experimental uncertainty of the $J_{1}=-0.09\pm0.03$ meV parametrization range. This quantum shifting is best illustrated in Fig.~\ref{fig:SQ}, where the structure factor calculated from classical simulations at $J_{1}=-0.0335$ meV is almost identical to quantum (NLC) results at $J_{1}=-0.06$ meV. As for neutron scattering measurements of Yb$_{2}$Ti$_{2}$O$_{7}$ (see Fig. 2.a of Ref.~[\onlinecite{Chang12a}]), the comparison is noticeably better with classical simulations for $J_{1}=-0.0335$ meV where a double transition takes place, than for $J_{1}=-0.09$ meV, confirming one more time the relevance of our theory to Yb$_{2}$Ti$_{2}$O$_{7}$.

Last but not least, our work brings Yb$_{2}$Ti$_{2}$O$_{7}$ as the missing link between Yb$_{2}$Sn$_{2}$O$_{7}$ and Yb$_{2}$Ge$_{2}$O$_{7}$, whose ground states are respectively splayed ferromagnetic~\cite{Yaouanc13a,Dun13a} and a yet not characterized antiferromagnet~\cite{Dun14a}, which we tentatively associate with U(1). On the basis of our work, we anticipate that a natural path will take this series of compounds through a transition from SFM to U(1), achieved either under chemical pressure (Sn$\rightarrow$Ti$\rightarrow$Ge) or hydrostatic pressure, in analogy with spin ice~\cite{Zhou12a} and  Tb$_{2}$Ti$_{2}$O$_{7}$~\cite{Mirebeau02a} experiments, respectively.\\


\textit{Conclusion --} Using a palette of complementary numerical methods, our work sets a theoretical benchmark common to both powder and single crystals of Yb$_{2}$Ti$_{2}$O$_{7}$. We have shown how the multi-step ordering recently observed in bulk measurements~\cite{Lhotel14a} naturally arises from the competition between a SFM phase and a U(1) manifold mediated by order-by-disorder selection. On a more fundamental level, the underlying ordering mechanism should be understood as the joint consequence of both quantum and thermal fluctuations. The $T=0$ frontier between the U(1) manifold and the splayed ferromagnet is shifted by quantum fluctuations, raising the interesting possibility to modulate frustration by tuning quantum fluctuations. In the general context of multiphase competition, order-by-disorder can be viewed as a free-energy optimization process which reinforces the stability of the degenerate phase it acts upon, by naturally selecting the subset of configurations with higher entropy and/or quantum zero-point fluctuations.

In light of the numerous models and phases supported by pyrochlores, ranging from spin liquids and spin ices to (partially) ordered phases~\cite{Savary12a,Gardner10a,Gingras14a,Brooks14a,Javanparast15a,Rau15a}, and subsequent boundaries between them~\cite{Yan13a,Chern10b,Mcclarty14b}, our present work is a paradigmatic example of why the properties of frustrated magnets should generically be understood as the sum of competing phases, rather than coming from a single controlling state. Some of these properties would indeed seem to be ``coming from nowhere'' in absence of a global phase diagram. We expect such competition between neighboring phases to be particularly relevant to some of the most difficult materials to characterize, such as Tb$_{2}$Ti$_{2}$O$_{7}$~\cite{Taniguchi13a} and Er$_{2}$Sn$_{2}$O$_{7}$~\cite{Sarte11a,Guitteny13b,Yan13a}, and to exacerbate the sample dependence issues~\cite{Ross12a,Taniguchi13a,Sala14a} by the proximity of phase boundaries. In that respect it is interesting to note that Er$_{2}$Ti$_{2}$O$_{7}$, whose coupling parameters lie far away from any phase boundary~\cite{Savary12b,Wong13a,Yan13a}, is one of the most robust rare-earth pyrochlore for reproducibility of experiments, while Yb$_{2}$Ti$_{2}$O$_{7}$ is essentially the antithesis. Hence, we expect the interplay between multiphase competition and disorder to become a very topical question, necessary to account for experiments in pyrochlores and frustrated magnetism, and promisingly rich in exotic physics.

\begin{acknowledgments}
The authors would like to thank Han Yan for collaboration on related projects. LDCJ, OB and NS are supported by the Okinawa Institute of Science and Technology Graduate University. RRPS is supported in part by NSF-DMR 1306048. The work at the U. of Waterloo was supported by the NSERC of Canada, the Canada Research Chair program (M.G., Tier 1) and by the Perimeter Institute (PI) for Theoretical Physics.  Research at PI is supported by the Government of Canada through Industry Canada and by the Province of Ontario through the Ministry of Economic Development \& Innovation.
\end{acknowledgments}


\section{Appendix}

\subsection{Order parameters and correlators}

As illustrated in Fig.~1 of the main text, three main phases are studied in this paper: the U(1) manifold, divided into $\psi_{2}$ and $\psi_{3}$ phases, and the splayed ferromagnet (SFM). Their order parameters can be found in the literature~\cite{Chern10b,Savary12b,Zhitomirsky12a,Yan13a,Mcclarty14b} and are repeated below for convenience.

In what follows, we adopt the convention of Ref.~[\onlinecite{Ross11a}] regarding the labeling of the spins, defined by their positions relative to the centre of the tetrahedron
\begin{eqnarray}
{\vec r}_0 = \left( \dfrac{1}{2},\dfrac{1}{2},\dfrac{1}{2} \right),&\quad
{\vec r}_1 = \left( \dfrac{1}{2},-\dfrac{1}{2},-\dfrac{1}{2} \right),\nonumber \\
{\vec r}_2 = \left(-\dfrac{1}{2},\dfrac{1}{2},-\dfrac{1}{2} \right),&\quad
{\vec r}_3 = \left(-\dfrac{1}{2},-\dfrac{1}{2},\dfrac{1}{2} \right).
\label{eq:r}
\end{eqnarray}
The spins at the four sites are denoted as $\vec S_{i=0,1,2,3}=(S_{i}^{x},S_{i}^{y},S_{i}^{z})$ in the cubic coordinates, with $|\vec S_{i}|=1/2$.

For the U(1) manifold, the order parameter, per tetrahedron, is a two-dimensional vector.
\begin{widetext}
\begin{eqnarray}
\vec m_{\rm U(1)}= \begin{pmatrix} m_{\rm U(1)}^{\alpha}\\ m_{\rm U(1)}^{\beta} \end{pmatrix}=
\begin{pmatrix}
\frac{1}{2 \sqrt{6} } \left( -2 S_0^x + S_0^y + S_0^z - 2 S_1^x - S_1^y-S_1^z+2 S_2^x + S_2^y-S_2^z +2 S_3^x-S_3^y +S_3^z \right) \\
\frac{1}{2 \sqrt{2}} \left( -S_0^y+S_0^z+S_1^y-S_1^z-S_2^y-S_2^z+S_3^y+S_3^z \right)
\end{pmatrix}
\label{eq:OPU1}
\end{eqnarray}
\end{widetext}
To differentiate between the $\psi_{2}$ and $\psi_{3}$ states within the U(1) manifold, we define the angle $\theta_{\rm U(1)}=\arctan(m_{\rm U(1)}^{\beta}/m_{\rm U(1)}^{\alpha})$; the function $\cos(6 \theta_{\rm U(1)})$ respectively equals to $+1$ and $-1$ for $\psi_{2}$ and $\psi_{3}$ states~\cite{Chern10b,Yan13a,Mcclarty14b}.\\

As for the splayed ferromagnetic phase, it is fully described by two three-dimensional order parameters~\cite{Yan13a}
%
\begin{eqnarray}
\vec m_{\rm SFM 1}&=& \begin{pmatrix} m_{\rm SFM1}^{\alpha}\\ m_{\rm SFM1}^{\beta}\\ m_{\rm SFM1}^{\gamma} \end{pmatrix}=
\begin{pmatrix}
\frac{1}{2} (S_0^x+S_1^x+S_2^x+S_3^x)\\
\frac{1}{2} (S_0^y+S_1^y+S_2^y+S_3^y)\\
\frac{1}{2} (S_0^z+S_1^z+S_2^z+S_3^z)
\end{pmatrix}
\end{eqnarray}

\begin{eqnarray}
\vec m_{\rm SFM 2}&=& \begin{pmatrix} m_{\rm SFM2}^{\alpha}\\ m_{\rm SFM2}^{\beta}\\ m_{\rm SFM2}^{\gamma} \end{pmatrix}\\
&=&
\begin{pmatrix}
\frac{-1}{2\sqrt{2}} (S_0^y+S_0^z-S_1^y-S_1^z-S_2^y+S_2^z+S_3^y-S_3^z)\\
\frac{-1}{2\sqrt{2}} (S_0^x+S_0^z-S_1^x+S_1^z-S_2^x-S_2^z+S_3^x-S_3^z)\\
\frac{-1}{2\sqrt{2}} (S_0^x+S_0^y-S_1^x+S_1^y+S_2^x-S_2^y-S_3^x-S_3^y) 
\end{pmatrix}\nonumber
\label{eq:OPSFM}
\end{eqnarray}
%
where $\vec m_{\rm SFM 1}$ is simply the global magnetization and $\vec m_{\rm SFM 2}$ accounts for the splayed nature of the ferromagnetism, \textit{i.e.} the fact that the spins are not colinear. For a given SFM ground state, both $\vec m_{\rm SFM 1}$ and $\vec m_{\rm SFM 2}$ are finite.

In order to compare the growth of correlations between the two phases, we define the order-parameter correlators
\begin{eqnarray}
C_{IJ}=\langle m_{I} m_{J}\rangle-\langle m_{I}\rangle\langle m_{J}\rangle
\label{eq:corr}
\end{eqnarray}
where $m_{I}$ is a given order parameter. For the U(1) manifold, the use of the correlators is rather straightforward. However, for the SFM phase, since both $\vec m_{\rm SFM 1}$ and $\vec m_{\rm SFM 2}$ are finite, one needs to compute the matrix of correlators
\begin{eqnarray}
\begin{pmatrix}
C_{\rm SFM1, SFM1} & C_{\rm SFM1, SFM2}\\
C_{\rm SFM2, SFM1} & C_{\rm SFM2, SFM2}
\end{pmatrix}.
\label{eq:matrix}
\end{eqnarray}
By definition, this matrix is symmetric but a priori non-diagonal. Upon diagonalization, the maximum eigenvalue is kept as correlator of the SFM phase. Please note that by symmetry of the lattice, calculations can be simplified by considering only one component of the order parameters, say \textit{e.g.} $\vec m_{\rm SFM1}^{\alpha}$ and $\vec m_{\rm SFM2}^{\alpha}$.

\subsection{Monte Carlo simulations}

The Fig.~1 of the main text has been obtained by classical Monte Carlo simulations of Heisenberg spins on the pyrochlore lattice. The conventional cubic unit cell consists of 16 spins and the system size has $N=3456$ spins. We used the standard Metropolis algorithm where a Monte Carlo step (MCs) is defined as $N$ random single-spin-flip attempts. To improve the quality of the simulations, parallel tempering~\cite{swendsen86a,geyer91a} and over-relaxation~\cite{creutz87a} were included in the simulations.

The error bars of Fig.~1 (main text) were obtained by using two different cooling procedures during the equilibration of the simulations. For the first ``annealed'' procedure, the initial configuration is chosen randomly; the system is then gradually cooled down from high temperature (fixed at 10 K) to the temperature of measurement $T$ during $10^{6}$ MCs; it is then equilibrated at the temperature $T$ during $10^{6}$ additional Monte Carlo steps.  Since it is starting from high temperature, the annealed procedure tends to favor the phase with higher entropy, \textit{i.e.} the U(1) manifold, and provides a lower boundary to the transition temperature. For the second ``quenched'' procedure, the initial configuration is fixed in the SFM phase; the system is then equilibrated at temperature $T$ during $10^{6}$ additional Monte Carlo steps. Since it starts in the ordered SFM phase, the quenched procedure favors the SFM phase and provides an upper boundary to the transition temperature. Following these equilibration procedures, measurements are done every 10 MCs during $10^{7}$ MCs.

For the phase diagram of Fig.~3 b) (main text), the system size was $3456$ sites with measurements during 10$^{6}$ MCs. For the structure factors of Fig.~3 a) (main text), no parallel tempering was necessary for simulations above the transition temperature. The system size was $128000$ sites with measurements during 10$^{6}$ MCs.

\subsection{Classical low-temperature expansion}
\label{sec:CSW}

Classical low-temperature expansion is an expansion in small fluctuations around an ordered state of classical spins. It enables calculation of the free energy of a given ordered phase up to leading term in temperature. We have used it to determine the low-temperature dependence of the phase boundary between the SFM and $\psi_3$ phases, for comparison with MC simulation as indicated by the green line in Fig. 1 of the main text. The method is a standard one, outlined in (e.g.) Ref. [\onlinecite{Shannon10a}].

\subsection{Linear spin wave theory}
\label{sec:LSW}

Linear spin wave theory (LSW) is a semi-classical method to study the stability of a classical phase in presence of the quantum zero-point energy. When the phase is a classical ground state, such as the SFM phase in the double-transition region of our paper, the method is rather straightforward. But if the phase is \textit{not} a classical ground state, such as the U(1) manifold in the double-transition region, the inclusion of zero-point energy requires a variation of the LSW theory, as proposed in Ref.~[\onlinecite{Coletta13a}] and outlined below.

At first, the approach is similar to standard spin wave expansion. We rewrite the spin operators ${\vec S}_i$ in terms of Holstein-Primakoff bosons and perform a $1/S$ expansion around the local spin configuration of the chosen ordered state. At the harmonic level, our Hamiltonian takes the form
$\mathcal{H} \approx \mathcal{H}_{\sf LSW}
=\mathcal{H}_{\sf LSW}^{(0)}+\mathcal{H}_{\sf LSW}^{(1)}+\mathcal{H}_{\sf LSW}^{(2)}$. $\mathcal{H}_{\sf LSW}^{(0)}$ is simply the classical energy of the ordered state around which we are expanding, while $\mathcal{H}_{\sf LSW}^{(1)}$ and $\mathcal{H}_{\sf LSW}^{(2)}$ contain only linear and quadratic terms in boson operators, respectively.

If the chosen ordered state is a classical ground state, then $\mathcal{H}_{\sf LSW}^{(1)}$ must vanish. As for $\mathcal{H}_{\sf LSW}^{(2)}$, it can be diagonalized by Fourier transform followed by a Bogoliubov transformation which will return real, positive, frequencies and therefore a meaningful excitation spectrum. The semi-classical energy can then be written
\begin{eqnarray}
E^0_{\sf semi-cl}=E^0_{\sf cl} \left(1+\frac{1}{S} \right) + \frac{1}{2} \sum_{\vec k \lambda} \omega_{\vec k \lambda},
\label{eq:Escl}
\end{eqnarray}
where $E^0_{\sf cl}$ is the classical ground state energy, $\omega_{\vec k \lambda}$ is the spin wave dispersion of band $\lambda$ at wave-vector $\vec k$.

However, if the chosen ordered state is \textit{not} a classical ground state, then $\mathcal{H}_{\sf LSW}^{(1)}$ may still vanish if the spin configuration we are expanding around is a \textit{saddle point} of the classical energy. As shown in Ref.~[\onlinecite{Savary12b}], this is the case for the $\psi_2$ and $\psi_3$ configurations. The last remaining point is the diagonalization of $\mathcal{H}_{\sf LSW}^{(2)}$. Following the approach developped by Coletta \textit{et al.}~\cite{Coletta13a}, we add to $\mathcal{H}_{\sf LSW}^{(2)}$ the positive definite term
$V=\delta \sum_i a_i^{\dagger} a_i$, where the $a_i^{\dagger}$ and $a_i$ operators are Holstein-Primakoff bosons. This additional term $V$ does not change the classical energy and the parameter $\delta$ may be adjusted to the minimum value for which we can obtain a real, positive excitation spectrum. The energy calculated with the inclusion of $V$ is
\begin{eqnarray}
E^{\delta}_{\sf semi-cl}=E_{\sf cl} \left(1 + \frac{1}{S} \right) - N \frac{\delta}{2}
+ \frac{1}{2} \sum_{\vec k \lambda} \omega_{\vec k \lambda}^{\delta}
\label{eq:Escl_delta}
\end{eqnarray} 
where $E_{\sf cl}$ is the classical energy of the saddle-point configuration and $\omega_{\vec k \lambda}^{\delta}$ is the spin wave dispersion calculated including the 
potential $V$. Since $V$ is a positive definite operator, $E^{\delta}_{\sf semi-cl}$ is an upper bound on the semiclassical energy of the saddle-point configuration.


\subsection{Exact diagonalization}
We consider the properties of the exchange Hamiltonian shown in Eq. (1) of the main text for a single tetrahedron ($N=4$) and a single cubic unit cell of the pyrochlore lattice ($N=16$) with periodic boundary conditions. Thanks to the small number of sites the full spectrum is analytically accessible for $N=4$ and the lowest lying eigenstates can be easily found via standard Lanczos diagonalization for $N=16$.


\subsection{Numerical linked cluster expansion}
At finite temperature, we have also performed a numerical linked cluster expansion (NLC)~\cite{Rigol07b,Tang13a}, defined as
\begin{equation}
P(\mathcal{L})/N = \sum_{\mathcal{C} \subset \mathcal{L}} L(\mathcal{C}) W(\mathcal{C})
\end{equation}
where $P$ is some extensive quantity and $N$ is the number of lattice sites. The
sum runs over clusters $\mathcal{C}$ of the lattice, $L(\mathcal{C})$ counts
the number of clusters of type $\mathcal{C}$ per site and $W(\mathcal{C})$ is the weight
evaluated on the cluster. This weight is computed using inclusion-exclusion rule

\begin{equation}
 W(\mathcal{C}) = P(\mathcal{C}) - \sum_{C' \subset C} W(\mathcal{C}')
\end{equation}
where $P(\mathcal{C})$ is the quantity computed on the cluster $\mathcal{C}$ and the sum runs over proper subclusters of $\mathcal{C}$. There is some freedom in choosing the classes of clusters to sum over in this expansion. We follow the approach of Ref.~[\onlinecite{Applegate12a}] and use tetrahedra as our building block. A linked cluster with $n_T$ tetrahedra will have at most $3n_T+1$ sites so we are limited to $n_T \leq 4$ in the expansion. The properties $P$ we will compute are defined on the tetrahedron and so conform well to this expansion. 
\begin{widetext}
\begin{center}
\begin{table}[h!]
\begin{tabular}{ c c m{6cm} cc m{6cm} }
\centering
$\mathcal{C}$ & $L(\mathcal{C})$ & &
$\mathcal{C}$ & $L(\mathcal{C})$ & \\
\hline 
$\mathcal{C}_0$ &
$1$  &
{\tiny $\bullet$}\vspace{0.2cm}&
&
&
\\
$\mathcal{C}_1$ &
$\dfrac{1}{2}$  &
\includegraphics[scale=0.4]{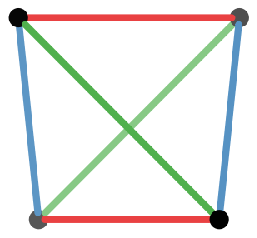} &
$\mathcal{C}_2$ &
$1$ &
\vspace{0.2cm} \includegraphics[scale=0.4]{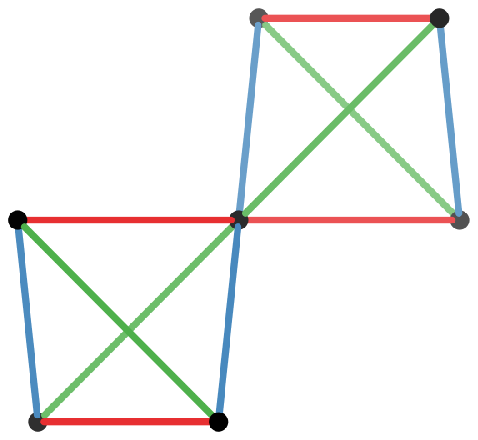} \\
$\mathcal{C}_3$ &
$3$ &
\includegraphics[scale=0.4]{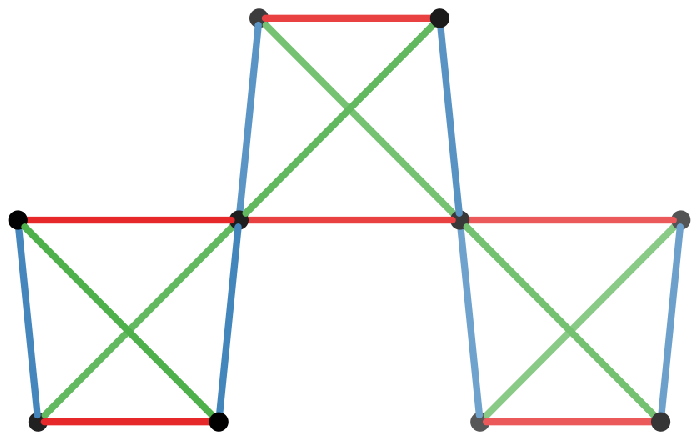} &
$\mathcal{C}_{4a}$ &
$3$ &
\includegraphics[scale=0.4]{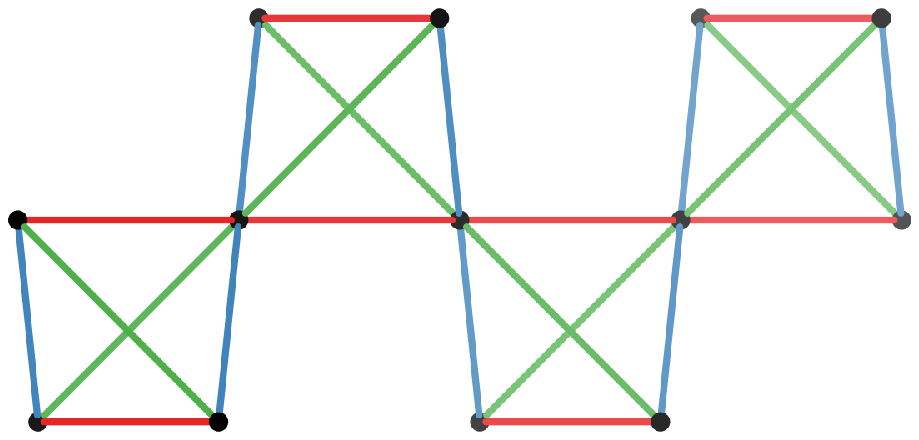} \\
$\mathcal{C}_{4b}$ &
$6$ &
\includegraphics[scale=0.4]{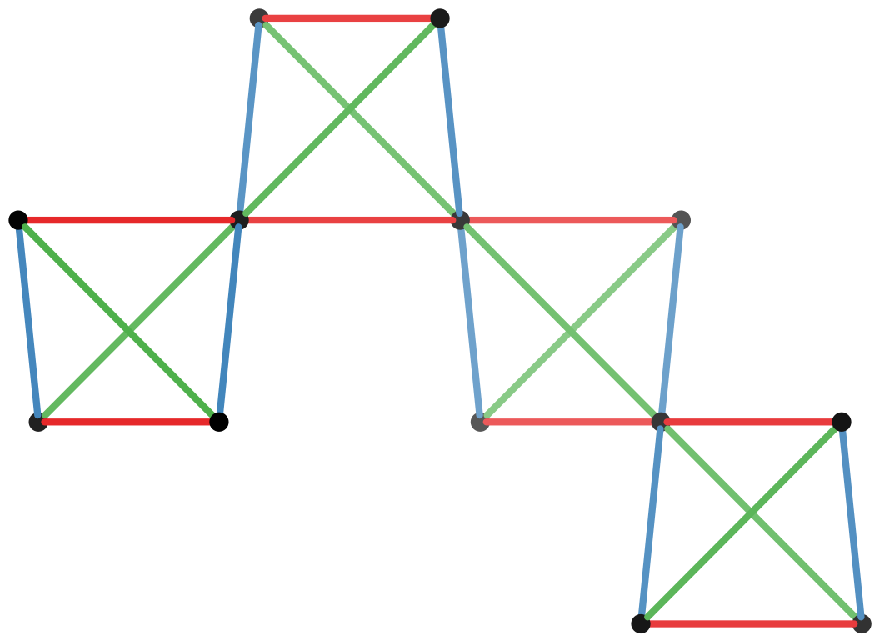} &
$\mathcal{C}_{4c}$ &
$2$ &
\includegraphics[scale=0.4]{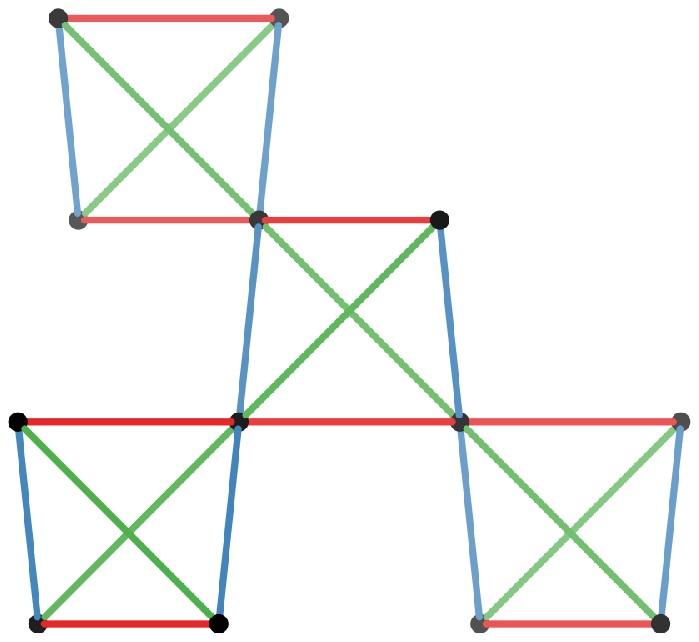} 
\end{tabular}
\caption{
\label{fig:clusters}
Clusters used for the NLC expansion. A graphical representation is shown along with the embedding constant $L(\mathcal{C})$.}
\end{table}
\end{center}
\end{widetext}

For convenience, we reproduce in Table \ref{fig:clusters} the geometrical clusters used in Ref.~[\onlinecite{Applegate12a}], with the appropriate embedding constant. These include the 0\textsuperscript{th} order point up to 4\textsuperscript{th} order which includes clusters composed of four tetrahedra.  Each cluster has an associated Hamiltonian $H_{\mathcal{C}}$ obtained from the exchange Hamiltonian (Eq. (1) of the main text) which we diagonalize numerically. The largest $n_T=4$ clusters we consider have 13 sites and thus Hilbert spaces of dimension $2^{13} = 8192$. These remain amenable to full diagonalization and thus we can compute arbitrary thermodynamic quantities at finite temperature.

The series is organized into terms $P_n$ including up to $n$ tetrahedra. Explicitly
carrying out the expansion using the embedding constants one has
\begin{eqnarray}
P_0 &=& +P(\mathcal{C}_0) \\
P_1 &=& -P(\mathcal{C}_0)+\frac{1}{2}P(\mathcal{C}_1) \\
P_2 &=& -\frac{3}{2} P(\mathcal{C}_1)+P(\mathcal{C}_2) \\
P_3 &=& +\frac{3}{2} P(\mathcal{C}_1)-5P(\mathcal{C}_2) + 3 P(\mathcal{C}_3) \\
P_4 &=& -\frac{1}{2} P(\mathcal{C}_1)+10 P(\mathcal{C}_2) - 21 P(\mathcal{C}_3) \nonumber\\
&&+ 3 P(\mathcal{C}_{4a}) + 6 P(\mathcal{C}_{4b}) + 2 P(\mathcal{C}_{4c})
\end{eqnarray}
Note that $P(\mathcal{C}_0)$ does not appear directly past first order. Following Ref.~[\onlinecite{Applegate12a}] the Euler resummation method is used on the final two terms to accelerate convergence. If we define the differences $S_n = P_{n+1}-P_n$ then the Euler approximants are given by
\begin{eqnarray}
  E_2 &=& P_2 \\
  E_3 &=& E_2 + \frac{1}{2} S_3 \\
  E_4 &=& E_3 + \frac{1}{4}\left(S_3+S_4\right)
\end{eqnarray}
The difference between 3$^{\rm rd}$ and 4$^{\rm th}$ order of expansion is the uncertainty of our NLC computations.
\subsection{High Temperature Expansion}

The High Temperature Expansions (HTE) are done up to order $\beta^{8}$, as explained in the book of Ref.~[\onlinecite{Oitmaa06a}]. The series expansion are then analyzed using Pad\'e approximants, \textit{i.e.} based on rational functions. We constructed all near-diagonal Pad\'e approximants with 8 or 7 terms in the series, \textit{i.e.} [4/4], [5/3], [3/5], [6/2], [2/6], [4/3], [3/4], [5/2], [2/5] where [m/n] stand for the powers of the polynomial in the numerator and the denominator. The resulting spread in Pad\'e approximant values represents the error bars of our HTE computations.

\bibliography{/Users/Ludo/Desktop/biblio}
\end{document}